\documentclass[aoas,MSNbibl,nameyear,seceqn,noautosecdot,dvips]{arximspdf}
\usepackage{stfloats}
\usepackage{graphicx}
\usepackage{dcolumn}

\doi{10.1214/10-AOAS440}
\volume{5}
\issue{2B}
\pubyear{2011}
\firstpage{1428}
\lastpage{1455}

\makeatletter
\fnbelowfloat
\newcolumntype{d}[1]{D{.}{.}{#1}}
\def\ci{\perp\!\!\!\perp}
\newcommand\Perp{\protect\mathpalette{\protect\independenT}{\perp}}
\def\independenT#1#2{\mathrel{\rlap{$#1#2$}\mkern4.1mu{#1#2}}}
\renewcommand{\ci}{\Perp}

\def\@bmisc[#1]{%
  \get@battribute{unstr}%
  \common@pub@types%
  \let\bauthor\bbl@bauthor%
  \let\bhowpublished\@firstofone%
  \def\borganization##1{{\bauthor@style ##1}}%
}

\makeatother

\begin{document}
\begin{frontmatter}

\title{Causal inference in transportation safety studies: Comparison of
potential outcomes and~causal~diagrams}
\runtitle{Causal inference in transportation safety studies}

\begin{aug}
\author[A]{\fnms{Vishesh}~\snm{Karwa}\ead[label=e1]{vishesh@psu.edu}},
\author[A]{\fnms{Aleksandra~B.}~\snm{Slavkovi\'{c}}\corref{}\ead[label=e2]{sesa@stat.psu.edu}}
\and
\author[B]{\fnms{Eric~T.}~\snm{Donnell}\ead[label=e3]{edonnell@engr.psu.edu}}
\runauthor{V. Karwa, A. B. Slavkovi\'{c} and E. T. Donnell}
\affiliation{Pennsylvania State University}
\address[A]{V. Karwa\\
A. B. Slavkovi\'{c}\\
Department of Statistics\\
Pennsylvania State University\\
University Park, Pennsylvania 16802\\USA\\
\printead{e1}\\
\phantom{\textsc{E-mail}: }\printead*{e2}} 
\address[B]{E. T. Donnell\\
Department of Civil and \\ \quad Environmental Engineering\\
Pennsylvania State University\\
University Park, Pennsylvania 16802\\USA\\
\printead{e3}}
\end{aug}

\received{\smonth{8} \syear{2009}}
\revised{\smonth{10} \syear{2010}}

%
\begin{abstract}
The research questions that motivate transportation safety studies are
causal in nature. Safety researchers typically use observational data
to answer such questions, but often without appropriate causal
inference methodology. The field of causal inference presents several
modeling frameworks for probing empirical data to assess causal
relations. This paper focuses on exploring the applicability of two
such modeling frameworks---Causal Diagrams and Potential Outcomes---for
a specific transportation safety problem. The causal effects of
pavement marking retroreflectivity on safety of a road segment were
estimated. More specifically, the results based on three different
implementations of these frameworks on a real data set were compared:
Inverse Propensity Score Weighting with regression adjustment and
Propensity Score Matching with regression adjustment versus Causal
Bayesian Network. The effect of increased pavement marking
retroreflectivity was generally found to reduce the probability of
target nighttime crashes. However, we found that the magnitude of the
causal effects estimated are sensitive to the method used and to the
assumptions being violated.
\end{abstract}

%

\end{frontmatter}

\section{\texorpdfstring{Introduction.}{Introduction}}\label{sec1}
An estimated 2.2 million people suffered some kind of
transportation-related injury in 2007. About 87 percent of these
injuries resulted from highway crashes [\citet{USDOT}]. Transportation
safety management aims at identifying causes of such crashes,
developing countermeasures to mitigate crashes, and evaluating the
effectiveness of a safety countermeasure. It is well known that causal
propositions of this kind, and their effect sizes, are best estimated
from randomized experiments.

The types of data available in transportation safety studies are
primarily observational, which makes it difficult to consistently
estimate causal effects of countermeasures. In this paper we evaluate
and compare the application of two commonly used causal inference
frameworks (one that is commonly applied in computer science and
another that is commonly applied in statistics) to transportation
safety. In particular, the aim of this paper is twofold:

\begin{itemize}
\item To introduce a unique transportation safety data set, created
from multiple sources, and to highlight the problems associated with
the data used in safety studies.
\item To explore the application of causal inference methods in
transportation safety studies and document the issues associated with
the analyses. We do this by estimating the causal effect of pavement
marking retroreflectivity ($\mathit{PMR}$) on target nighttime crashes using the
causal modeling frameworks of the Potential Outcomes (PO) and Causal
Diagrams (CD), and then compare the results.
\end{itemize}

Causal inference methods in transportation safety studies have received
little attention. \citet{Davis2000} provides a review and notes that
the assignment mechanism must be included in statistical models to
consistently estimate the effect of any countermeasure on crashes.
\citet{Davis2004} uses Pearl's causal Bayesian networks (CBN) for crash
reconstruction and examines token causal claims to answer single-event
causation questions; see \citet{Eells91} for a review of token and type
causes.\footnote{Token causes are those associated with a single unit or
event, for instance, ``Since Jane was speeding, she ended up in an
accident.'' On the other hand, type causal claims are associated with a
population, for instance, ``Speeding causes crashes.''} In contrast to
single-event causation, our work examines the application of causal
inference methods to population level causal effects in transportation
safety studies. Population causal claims are more applicable to
transportation safety management since they reflect the effect of
countermeasures in a population as opposed to singular causal claims
which are geared more toward accident reconstruction and liability
issues. We examine the hypothesis that low $\mathit{PMR}$ levels are causative
agents for an increase in the risk of nighttime crashes. Causal effects
are estimated using the PO framework and CD framework, and the results
are compared.

It has been shown that the PO framework and the CD framework are
mathematically (theoretically) equivalent; see \citet{Pearl00causalitymodels}, Chapter~7. However, there are different statistical
implementations of these frameworks that offer different paths to
estimate causal effects, and, in practice, the results may or may not
be similar. For instance, the PO framework is commonly implemented
using propensity score matching or inverse propensity score weighting,
and the CD framework is commonly implemented using CBNs. Several
assumptions relating to estimation of a causal effect from
observational data are reviewed in Sections \ref{sec:po} and \ref
{sec:cd}. Apart from these, differences in the estimates of causal
effect could arise due to additional assumptions required by each
framework, inherently tied to the aforementioned statistical
implementations. For instance, the CD framework requires the use of a~%
causal graph that represents the qualitative causal mechanism of the
data generating process. This graph can be obtained from prior
knowledge and/or data. The algorithms used to recover causal graphs
from data require additional assumptions such as faithfulness and some
require the data to be either discrete or Gaussian. The PO framework
does not require these additional assumptions since it does not require
such causal graphs.\footnote{The PO framework requires the analyst to
qualitatively model the treatment assignment mechanism (see Section~\ref{sec4.1}). The use of graphs to represent the treatment assignment mechanism
could make this easier to communicate; see the discussion section.}
Furthermore, while the CD framework enables one to work with the
complete data set, the PO framework could lead to elimination of a part
of the data set if one uses matching.
\citet{sfer} and \citet{fiensfer} show in a~simple simulated logistic
regression example that there is an implicit agreement between PO and
CD frameworks. However, we are not aware of a~study that explores the
differences, compares the results of both modeling frameworks on
real-life observational data, and examines the advantages and
disadvantages associated with applying each method in a practical context.

A complete and rigorous comparison of both frameworks requires
considerations of all possible implementations, which is beyond the
scope of this paper. Here, we focus on two commonly used
implementations in practice, both of which use the complete data set in
order to allow for a better comparison of the results. We implement the
PO framework using \textit{inverse propensity score weighting} (\textit{IPW}) and
\textit{regression adjustment}, and the CD framework using \textit{discrete causal
Bayesian Networks} (CBN); the implementation details are provided in
Sections~\ref{sec:po} and \ref{sec:cd}. We also implemented the PO
framework using \textit{propensity score matching}\footnote{We report the
estimates of causal effects from matching in the paper, and provide the
implementation details in the supplementary material.} since it is a
popular alternative to IPW. However, as previously noted, matching
could lead to elimination of part of the data and the comparison of
estimates with those from CBN may be biased due to differences in the
data themselves.
This issue is further discussed in Sections \ref{sec:po} and \ref
{con.dis} of the paper.

The remainder of the paper is organized as follows: Section \ref
{sec:problem.des} introduces the problem statement; Section \ref
{sec:design} introduces the data set and the design of the hypothetical
experiment to estimate the causal effects; Sections \ref{sec:po} and
\ref{sec:cd} present the analyses and results of the PO framework and
the CD framework; Section \ref{sec:comp.res} compares the results; and,
Section \ref{con.dis} concludes with a~discussion.

\section{\texorpdfstring{Problem description.}{Problem description}}
\label{sec:problem.des}
The traffic accident fatality rate increases by almost 75 percent
during the period of time between 9 p.m. and 6 a.m. [\citet{NHSTA}].
This raises a question of what can be done about the fact that there is
a greater rate of crashes at night than during the day?
It is hypothesized that low pavement marking visibility may be one
cause of the increased rate of nighttime crashes. One of the objectives
of the paper is to examine this hypothesis.

Pavement markings delineate the limits of the traveled way and provide
drivers navigation and control guidance. During the daytime, drivers
are likely to use a combination of pavement markings, other traffic
control devices (e.g., signs) and visual cues along the roadside (e.g.,
utility poles, vegetation, etc.) to navigate a roadway. Pavement
markings have an important role at night. Apart from delineating the
road, pavement markings reflect the light shone from a car's headlamps
back to the driver, thus enabling the driver to see the limits of the
traveled way. This is known as retroreflectivity and is measured in
millicandelas per square meter per lux (mcd/m$^2$/lux).
Retroreflectivity in pavement markings is provided by glass spheres
that are dropped-on or premixed with a wet pavement marking material.
$\mathit{PMR}$ degrades over time because of fatigue to the material and its
bond strength with glass spheres or the pavement surface. State
transportation agencies typically re-stripe pavement markings after the
end of their useful service life, defined as the time when
retroreflectivity falls below a minimum threshold level.
%

A considerable amount of research has been carried out regarding the
safety benefits of $\mathit{PMR}$. To answer the question if improving $\mathit{PMR}$ has
any effect in reducing the number of traffic crashes, most of the
published literature used regression models with observational data,
ignoring the treatment assignment mechanisms; for a literature review,
refer to \citet{bahar} and \citet{donnell}. Also \citet{donnell} point
out that none of the studies explicitly relate the in-situ $\mathit{PMR}$ levels
to the crash event. This is due to the fact that $\mathit{PMR}$ levels and crash
data are obtained from separate sources and merging them is difficult.
The problems associated with merging these databases have been
described in \citet{visheshthesis}.
\citet{donnell} was the first study that explicitly combined the $\mathit{PMR}$
data (which is representative of real life degradation patterns of
$\mathit{PMR}$) with crash data to develop a comprehensive database. This work
did show that there were statistical associations between PMR and
nighttime crashes. We use this database to examine, for the first time,
the nature of the effect of $\mathit{PMR}$ on traffic safety (defined in Section
\ref{sec:design}) using the PO and the CD frameworks. The results are
compared with a discussion of the application of the two methods, in an
attempt to determine their possible broader application in
transportation safety studies.

\section{\texorpdfstring{Description of the data and design of the study.}{Description of the data and design of the study}}
\label{sec:design}
The fundamental unit of operation in this paper is a homogeneous road
segment; homogeneous refers to having uniform geometric characteristics
such as number of lanes, lane width and shoulder width along a roadway
segment. A segment within a fixed time period is considered to be
different from the same segment at any other time period. A fixed time
period of one month was selected to ensure homogeneity of $\mathit{PMR}$ levels
and other characteristics of a segment. For instance, the PMR level and
monthly traffic volumes can be assumed to be reasonably uniform within
this period.

Crash and $\mathit{PMR}$ data were collected from three districts in North
Carolina for a period of 2.5 years. As noted in Section \ref
{sec:problem.des}, the data were obtained from two different sources.
The PMR data were measured by a private contractor using a mobile
retroreflectometer with a 30-meter geometry. These data were collected
on two-lane and multi-lane highways in North Carolina, approximately
every 6 months. All pavement markings were of thermoplastic material.
Since retroreflectivity estimates were not measured at the exact time
and place of occurrence of the crash, a neural network model was used
to interpolate the values of retroreflectivity on the segments where
crashes were observed; see \citet{karwa}.

The roadway inventory and crash event data were obtained from the
Highway Safety Information System (HSIS) data files, maintained by the
Federal Highway Administration (FHWA). These data were collected for 19
roadway sections in North Carolina. There were 192 total segments
(segments are a~subset of sections) that corresponded to the 19
sections of roadway where PMR estimates were computed based on the
degradation model. Table \ref{roadinfo} shows the sections where
roadway inventory, crash and PMR data could be linked. There are a
total of 5,916 observations, based on 192 segments, 12 months of data
per year for each segment and approximately 2.5 years of crash data per
segment.\footnote{The assumption of independent segments over time is a
common assumption in the safety prediction literature that also shows
that weak temporal (or spatial) correlations result in a loss of
estimation efficiency but not bias.}

%
\begin{table}
\tabcolsep=0pt
\caption{Roadways with pavement marking and crash data available for
safety analysis}
\label{roadinfo}
\begin{tabular*}{\textwidth}{@{\extracolsep{\fill}}lcd{2.0}d{2.2}d{2.2}cd{3.3}d{3.0}@{}}
\hline
  &  &  &  &
&  & \multicolumn{1}{c}{\textbf{Total}} & \multicolumn
{1}{c@{}}{\textbf{Nighttime}}\\
& & &  \multicolumn{1}{c}{\textbf{Begin}}& \multicolumn{1}{c}{\textbf{End}} &\multicolumn{1}{c}{\textbf{Number of}}
& \multicolumn{1}{c}{\textbf{length}} & \multicolumn{1}{c@{}}{\textbf{target}} \\
\textbf{County} &\multicolumn{1}{c}{\textbf{Route}} &\multicolumn
{1}{c}{\textbf{District}} & \multicolumn{1}{c}{\textbf{MP}}& \multicolumn{1}{c}{\textbf{MP}}&  \multicolumn{1}{c}{\textbf{lanes}}& \multicolumn{1}{c}{\textbf{(miles)}} & \multicolumn{1}{c@{}}{\textbf{crashes}}\\
\hline
Bertie & US13 & 1 & 0.00 & 11.07 & 4 & 11.07 & 8\\
Gates & US13 & 1 & 0.00 & 14.78 & 2 & 14.78 & 14\\
Northampton & US158 & 1 & 12.35 & 24.04 & 2 & 11.69 & 4\\
Washington & US64 & 1 & 10.54 & 19.67 & 2 & 9.13 & 2\\
Durham & I-85 & 5 & 7.88 & 14.19 & 4 & 6.31 & 5\\
Durham & US15 & 5 & 3.66 & 6.56 & 4 & 2.90 & 9\\
Durham & NC98 & 5 & 0.00 & 11.06 & 2 & 9.44\tabnoteref{a} & 15\\
Durham & NC157 & 5 & 0.70 & 3.98 & 2 & 3.28 & 2\\
Granville & I-85 & 5 & 0.00 & 23.73 & 4 & 1.80\tabnoteref{b} & 5\\
Person & US158 & 5 & 0.00 & 22.36 & 2 & 16.22 & 5\\
Vance & I-85 & 5 & 0.00 & 14.47 & 4 & 12.47\tabnoteref{c} & 46\\
Vance & US158 & 5 & 0.00 & 8.96 & 2 & 5.94\tabnoteref{d} & 1\\
Wake & I-40 & 5 & 6.47 & 20.19 & 4/6/8 & 13.72 & 67\\
Wake & NC98 & 5 & 0.00 & 4.55 & 2 & 4.55 & 1\\
Warren & I-85 & 5 & 0.00 & 9.88 & 4 & 9.88 & 39\\
Warren & US158 & 5 & 12.38 & 22.93 & 2 & 10.55 & 0\\
Catawba & I-40 & 12 & 13.13 & 19.67 & 4 & 6.54 & 10\\
Iredell & I-40 & 12 & 0.00 & 22.76 & 4 & 22.76 & 63\\
Iredell & I-77 & 12 & 14.75 & 23.75 & 4 & 9.00 & 17\\
[3pt]
Total & & & & & & 182.03 & 313\\
\hline
\end{tabular*}
\tabnotetext[a]{a}{Roadway inventory and crash data were not available between mileposts 0.17 \&
1.79.}
\tabnotetext[b]{b}{Roadway inventory and crash data were not available between mileposts 0.76 \& 22.69 and
22.73.}
\tabnotetext[c]{c}{Roadway inventory and crash data were not available between mileposts 3.96 and
5.96.}
\tabnotetext[d]{d}{Roadway inventory and crash data were not available between mileposts 3.77 and
6.79.}
\end{table}

Crashes that satisfied the following criteria, referred to as \textit{target crashes}, were used in the analysis: occurred during dusk, dawn
or at night; dry roadway surface conditions; ran-off-the-road crashes;
fixed object crashes (off-road); and opposite- or same-direction
sideswipe crashes. Crashes that satisfied the following criteria were
excluded from the analysis: work zone area; no alcohol-involvement;
weather contributing circumstances; roadway contributing circumstances.
It must also be noted that the data are sparse due to rarity of
crashes. Only $6$ percent of the total number of segments had more than
one target crash during the study period.

\textit{Safety} of a road segment is defined as a $\mathit{Bernoulli}$ random
variable, taking the value 1 if there was at least one target crash in
the segment during the treatment (or control) application period, and 0
if there was no target crash in the segment. The safety of a segment is
stochastic and each segment has a fixed probability $p$ of at least one
target crash occurring, which is assumed to be an inherent property of
the road segment.
This definition was chosen to ensure the absence of confounders between
safety and $\mathit{PMR}$, based on the past $\mathit{PMR}$ related safety literature.
For instance, if it was clear according to the police crash report that
a particular crash occurred due to driving under the influence of drugs
or alcohol, such a crash would have been deemed to occur because of
human error, and thus excluded from the current analyses. Similarly,
crashes in which weather was a contributing factor (such as heavy snow
or icy road conditions) were also excluded from the analyses. Weather
conditions, human errors, etc. are stochastic factors that may cause
crashes but not an inherent property of the segment; hence, any crash
occurring due to such conditions would fall into the error term of the
observed safety of a~road segment.

\textit{Treatment} variable on a segment is defined as the application of
$\mathit{PMR}$ with levels $ \{ \mathit{Low}, \mathit{Med}, \mathit{High} \}$; the exact range of $\mathit{PMR}$
levels for each class is specified in Table \ref{define}. \textit{Control}
is defined as application of pavement markings at one of the two
remaining levels of retroreflectivity.
Out of the total sample size ($N=5\mbox{,}916$), about 36 percent of the
segments had $\mathit{Low}$ levels of $\mathit{PMR}$, about 46.5 percent had $\mathit{Med}$ levels
and the remaining segments had $\mathit{High}$ levels of $\mathit{PMR}$. The assignment
of $\mathit{PMR}$ levels is clearly not random.

Apart from the data on $\mathit{PMR}$ and the crash counts per month, data on 12
other covariates were collected. See Table~\ref{define} for definitions
and summary statistics of random variables representing information on
the attributes of a segment such as the shoulder width, number of
lanes, presence of a~median, traffic flow characteristics such as
monthly traffic volumes (hereafter referred to as $\mathit{ADT}$), percentage of
trucks, location related variables such as the geographic district in
which the segment is located, the urban or rural setting of the segment
location, and the terrain type. The data are very sparse, which is
typical of safety data. For instance, in the five way
cross-classification of the entire sample with respect to the discrete
variables $\mathit{District}, \mathit{Terrain}, \mathit{PMR}, \mathit{Multilane}$ and $\mathit{Safety}$, $58$
percent of the cells have sampling zeros.

\begin{table}
\tabcolsep=0pt
\caption{Definition of variables and their descriptive statistics. Mean
(st. dev), [Min, Max] values are given for the continuous variables and
number of observations (percentage) for each level of categorical~variables. Total sample size, $N=5\mbox{,}916$}
\label{define}
\begin{tabular*}{\textwidth}{@{\extracolsep{\fill}}llc@{}}
\hline
\textbf{Variable}& \multicolumn{1}{c}{\textbf{Definition}} & \textbf{Descriptive statistics}\\
\hline
Right & Outer shoulder width in feet on right & 9.39 (4.03) \\
Shoulder&side of roadway& [0, 14]\\
& 0 if shoulder width $\leq$ 7 feet &1,238 (9.97\%) \\
& 1 otherwise & 4,678 (90.03\%)\\
[4pt]
ADT& Annual average daily traffic adjusted & 30,383 (27,580) \\
&for a month, vehicles per day& [1,615, 114,400]\\
&0 if ADT $\leq$ 30,000 vehicles per day&2,957 (49.9)\\
&1 otherwise& 2,959 (50.1)\\
[4pt]
Truck& Percentage of ADT that consists & 14.8 (8.5) \\
& of heavy vehicles& [0, 83]\\
&0 if Percentage of Trucks $\leq$ 18& 2,065 (35)\\
&1 otherwise& 3,851 (65)\\
[4pt]
PMR& Mean PMR of all markings on a segment& 227 (65) \\
& (mcd/m$^2$/lux)& [139, 447]\\
& Low if $139<$ retroreflectivity $\leq200$ &2,134 (36)\\
&Medium if $200<$ retoreflectivity $\leq280$ &2,748 (46.5)\\
&High if $280<$ retroreflectivity $\leq447$ &1,034 (17.5)\\
[4pt]
Age& Time (in months) elapsed since the application & 15.5 (8.63)\\
&of markings on a segment&[1, 30]\\
&0 if Age $\leq$ 1&1,761 (30)\\
&1 if 10 $\leq$ Age $\leq$ 20&1,980 (33.5)\\
&2 otherwise &2,175 (36.5)\\
[4pt]
Multilane& 1 if there is more than 1 lane in each direction& 3,868 (65.4)
\\
& 0 if there is 1 lane in each direction& 2,048 (34.6) \\
[3pt]
Median & 1 if the segment contains a median&4,018 (68) \\
& 0 if the segment has no median&1,898 (32)\\
[4pt]
Safety & 1 if at least 1 target crash occurred & 376 (6.36)\\
&in the segment, during the month &\\
& 0 otherwise& 5,540 (93.65)\\
[4pt]
Urban & 1 if the segment is located in an urban area& 2,680 (45.3) \\
& 0 if the segment is located in a rural area&3,236 (54.7)\\
[3pt]
Terrain & 1 if the segment is on flat terrain& 1,320 (22.3)\\
& 0 if the segment is on a rolling terrain&4,596 (77.7)\\
[4pt]
District &0 if the segment is located in District 1&1,290 (21.8)\\
& 1 if the segment is located in District 5&3,286 (55.5)\\
& 2 if the segment is located in District 12&1,340 (22.7)\\
\hline
\end{tabular*}
\end{table}

As per \citet{rubin-2008-2} and \citet{MaldonandoandGreenland2002}, we
conceptualize our problem as a hypothetical experiment to make the
problem statement clear. Consider a population of homogeneous road
segments. We wish to examine the effect of increased $\mathit{PMR}$ on the
safety of a road segment. Ideally, we would like to apply treatment
(e.g., $\mathit{PMR} = \mathit{Low}$) and control (e.g., $\mathit{PMR} = \mathit{High}$) to the same
population and observe the expected safety outcome to measure the
causal effect.\footnote{In practice, ``treatment'' would be application
of Higher PMR, but here, we define it as application of Low PMR to
obtain causal risk ratios greater than 1.} The causal effect is
defined as the risk ratio of expected safety outcome under the
treatment and controls, for the same population. Since this is not
possible in practice, we use analytical simulations of this process.
Sections~\ref{sec:po} and~\ref{sec:cd} describe the conceptualization of this hypothetical
experiment under two different frameworks.

\section{\texorpdfstring{Potential outcomes framework.}{Potential outcomes framework}}
\label{sec:po}
In this section we present the PO framework as applicable to the
current study as well as the results of the analysis. Section~\ref{1}
defines the causal estimands [the ``science,'' see \citet{Rubin2005}]
and \mbox{explains} the treatment assignment mechanism and the assumptions
required to estimate causal effects from observational data. Section
\ref{2} provides details about the implementation of the PO framework,
that is, the use of inverse propensity score weighting to achieve
balance in the data and the use of regression adjustment to estimate
the average causal effect (ACE), after balancing. Section \ref{3}
presents the results of the analysis.

\vspace*{-3pt}
\subsection{\texorpdfstring{Treatment assignment, potential outcomes and assumptions.}{Treatment assignment, potential outcomes and assumptions}}
\label{1}\label{sec4.1}
Let the homogeneous segments be indexed by the letter $i$. We focus on
one hypothetical experiment at a time, introduced in Section \ref
{sec:design}, and estimate the effect of a binary $\mathit{PMR}$ treatment on
safety from a sample of segments. Extension to the case of three levels
of treatment of $\mathit{PMR}$ is performed using the method proposed by \citet
{rubin1998} which involves creating a separate propensity score model
for each two-level treatment comparison, equivalent to conducting three
hypothetical experiments. Thus, in the present case, three separate
propensity score models are estimated. This method is followed since it
is difficult to simultaneously balance all three treatment groups on
all covariates.

The PO framework uses potential outcomes as the fundamental element to
estimate the causal effects. We denote the treatment variable by $
T_{i},$ where $ T_{i} = 0 $ denotes no treatment or the baseline
condition for unit $ i $, and $ T_{i} = 1 $ denotes the treatment
condition. For instance, if we wish to estimate the effect of changing
the $\mathit{PMR}$ levels from $\mathit{Med}$ to $\mathit{Low}$, the treatment would be
application of $\mathit{PMR}=\mathit{Low}$ and the control would be application of
$\mathit{PMR}=\mathit{Med}$. Associated with each segment are two potential outcomes:
$\mathit{Safety} (S) $ of the segment at the end of a month after the treatment
has been applied, $ S_{i}(T = 1) $, and $\mathit{Safety}$ of the same segment at
the end of the month if there was no treatment, that is, the baseline
condition was applied, $ S_{i}(T = 0)$. Covariates that represent the
attributes of a segment are denoted by the vector $ X_{i} = (x_{i1},
x_{i2}, \ldots  , x_{ip}) $ for unit $i$. The average causal effect
(ACE) of the treatment relative to the baseline for segment $i$ is then
defined as a~causal risk ratio\vspace*{-2pt}
%
\begin{equation}\label{riskratio}
\mathit{ACE}_{\mathrm{0\ to\ 1}} = \frac{E[S_i(1)]}{E[S_i(0)]},
\vspace*{-2pt}
\end{equation}
where $ E[\cdot] $ denotes expectation and $E[S_i(\cdot)] = E[E[S_i(\cdot)|P(S_i(\cdot)
= 1)]]$. We assume that $ E[S_i(0)] > 0 $, and drop the $T $ in the
notation for simplicity.

For any particular segment, only one of the two values of $S(0)$ and
$S(1)$ can be observed. This has been termed the ``fundamental problem
of causal inference'' [\citet{Rubin1978}; \citet{Holland86statisticsand}],
because of which unit level causal inferences are not possible.\footnote
{Except in cases where the functional mechanism of causation is known,
this is called token causation; see \citet{Pearl00causalitymodels}, Chapter 7.} However, given certain assumptions which
are outlined below, the ACE of the treatment on a population can be
estimated consistently.

\textit{Stable unit treatment value assumption} (SUTVA) [\citet
{Rubin1990}]. This assumption states that the treatment applied to one
unit does not affect the outcome of any other unit and that there are
no hidden versions of the treatment (i.e., no matter what mechanism was
used to apply the treatment to the unit, the outcome would be the
same). The last part is sometimes referred to as the consistency
assumption [\citet{coleconsistency}]. We make this assumption in the
current study even though the treatment has been applied in groups
(several segments along a particular route may have the same value of
$\mathit{PMR}$). The following example illustrates a scenario where this
assumption could be violated. Consider two consecutive segments on the
same route. A vehicle traveling on this road could end up in a crash in
segment 2 because of low visibility on segment 1. Such scenarios are
not uncommon, but when crashes are reported, the reporting officer
estimates the approximate segment location where the crash was
initiated (after careful analysis of the evidence available at the
crash site, such as skid marks, etc.) and the crash is attributed to
that segment. \citet{GuangleiHong01012005} extend SUTVA to account for
possible interference among segments, but we do not consider this
extension here.

\textit{Positivity.} The positivity assumption states that there is a
nonzero probability of receiving every level of treatment for every
combination of values of exposure and covariates that occur among
individuals in the population [\citet{Rubin1978}; \citet{HernanRobins}]. We
make the positivity assumption since, in principle, each segment can be
assigned any level of $\mathit{PMR}$ treatment.

\textit{Unconfoundness.} The treatment mechanism is said to be
unconfounded given a set of covariates $x_i$, if the treatment is
conditionally independent of the potential outcomes given the covariates
%
\begin{equation}
t_{i} \ci S(0),S(1) \mid x_{i}.
\end{equation}
In a randomized experimental setting, $t_{i}$ would be unconditionally
independent of the potential outcomes by design. In the current setting
this is not the case, but the treatment assignment can be made
conditionally independent of the potential outcomes by balancing on
observed covariates. This requires modeling the treatment assignment
mechanism as explained below.

\subsubsection{\texorpdfstring{Treatment assignment mechanism.}{Treatment assignment mechanism}}
Let $P(T_{i} = 1 \mid X_{i}) $ be the pro\-pensity score. The propensity
scores are used in assignment of the treatment to the segments in order
to achieve balance. Following \citet{roserubin1983}, treatment
assignment is strongly ignorable given a vector of covariates $X$ if
unconfoundedness and common overlap hold:
%
\begin{eqnarray}
&\displaystyle S(0),S(1) \ci T \mid X,& \\
&\displaystyle 0 < P(T=1 \mid X) < 1.&
\end{eqnarray}

In the current setting, the treatment assignment mechanism can be
assumed to consist of two parts. In the first part, pavement markings
are applied by transportation agencies at different segments with a
similar level of retroreflectivity (usually falling into the category
$\mathit{High}$). In the second part, the markings are left to deteriorate over
a period of 2.5 years. The $\mathit{PMR}$ levels decrease due to stress on the
pavement marking material from vehicle passes and natural factors such
as weather. Thus, it can be assumed that nature assigns a level of
$\mathit{PMR}$ based on the time elapsed since the initial application period
($\mathit{AGE}$) of the pavement marking, number of vehicle passes and weather
conditions. The assignment of $\mathit{PMR}$ levels for each segment depends on
the $\mathit{AGE}$ of the marking within the segment and the number of vehicle
passages over the segment within that period.
Apart from this, $\mathit{PMR}$ levels may also depend on the location of the
segment (due to differences in weather conditions), the percentage of
trucks that compose the traffic volumes (stress on the marking material
is generally greater due to heavier vehicles) and the number of lanes
in a segment (the $\mathit{PMR}$ levels used are the average of the different
pavement markings present in a segment, and multi-lane segments
generally have at least one extra marking when compared to two lane
segments). All of these variables are included to form a rich
propensity score model that specifies the assignment mechanism.

\subsection{\texorpdfstring{Inverse propensity score weighting and regression
adjustment.}{Inverse propensity score weighting and regression
adjustment}}\label{2}

Below is a description of a particular
implementation of the PO framework that estimates ACE and ensures that
the assumptions outlined in the previous sections are satisfied. This
implementation is a form of doubly robust estimation; see \citet
{bang2005doubly}. The same two steps are repeated to estimate the ACE
for each of the three comparisons:
\begin{itemize}[Step 2]
\item[Step 1] Estimate the propensity score model $\pi=P(T=1\mid X)$ and
achieve balance, via \textit{inverse propensity score weighting} (\textit{IPW});
see \citet{hirano2001}.
\item[Step 2] Estimate ACE, via \textit{regression adjustment method}.
\end{itemize}

\textit{Inverse propensity score weighting}: The Generalized Boosting
Model\break [GBM, \citet{mccaffrey2004propensity}], a multivariate
nonparametric technique, was used to estimate the propensity scores.
Although logistic regression is the most common way to estimate the
propensity scores, studies have shown that other methods can offer
considerable improvement [e.g., see \citet{lee2009improving}].
The analysis was carried out using the ``twang'' package in R [\citet
{ridgeway2006toolkit}].
Weights were computed from the estimated propensity scores and balance
in the data is tested using the estimated weights.
Balance is tested by comparing the distributions of key covariates in
the treatment and control groups of the weighted data using the
Kolmogorov--Smirnov (KS) test statistic. A weight of $\frac{1}{\pi}$ is
assigned to the treatment group and a weight of $\frac{1}{1-\pi}$ to
the control group, where $\pi$ is the estimated propensity
score.\footnote{These correspond to the population weights (ATE weights
in the twang package).} \citet{hirano2003efficient} show that the use
of a nonparametric estimate of propensity score to estimate weights,
rather than the true propensity score, can lead to an efficient
estimate of ACE. If balance is not achieved, the propensity score model
is re-specified and the process is repeated. The model is re-specified
by changing the tuning parameters of the boosting model. The tuning
parameters are the number of trees used to fit the model, the shrinkage
parameter and the interaction depth; see \citet
{mccaffrey2004propensity} for details. In the selected propensity score
models, we used an interaction depth of 2 (i.e., the model fits all
two-way interactions), the shrinkage parameter was set at 0.01 and the
number of trees were set at 15,000.

\textit{Regression adjustment}: Once the samples (segments) are divided
into control and treatment groups and balance is achieved, several
methods exist to estimate the ACE of the treatment [\citet
{Schaferkang2008}]. We applied the inverse propensity weighted
regression adjustment. In this method, a~model for the safety outcomes
(henceforth referred to as \textit{the outcome model}) under both
treatment and control application is estimated\footnote{One could also
specify two separate outcome models, one for the outcome under
treatment and one for the outcome under control. We follow this
approach in the matching implementation, which is described in the
supplementary material [\citet{supp}].} using weighted regression; the
weights come from the estimated propensity scores. We again make use of
GBM to estimate a single regression model by using an indicator for the
treatment. All covariates listed in Figure~\ref{fig:balance} are
included in the model, and we choose the interactions implicitly.
Overfitting is avoided by using 10-fold cross-validation and out-of-bag
estimation [\citet{ridgewaygbm}]. The GBM model of the outcomes is used
to predict the probability distributions of $S_{i} (1) $ and $ S_{i}(0)$.

Once the model for the outcomes is specified, we used two approaches to
estimate the ACE that is the causal risk ratio of equation~(\ref
{riskratio}). The standard errors for these estimates were obtained by
bootstrapping. In the \textit{combined prediction} approach, the complete
data are used to predict the outcome under treatment application using
the outcome model, and the complete data are used to predict the
outcome under control application. In the \textit{individual prediction}
approach, only treated data are used to predict the outcome under
treatment application, and only control data are used to predict the
outcome under the control application. In other words, in the combined
method, the complete data are used to estimate the potential outcomes
under treatment and control application, irrespective of the actual
assignment,\footnote{It must be noted that the combined prediction is
similar in spirit to the ``do'' operator, which will be introduced in
the Causal Diagrams section; see also the discussion section. Also, the
individual prediction can be considered to be an ``analytical
simulation'' of a~randomized experiment.} whereas in the individual
method the potential outcomes under treatment and control are estimated
by using that part of the data which actually received the treatment
and control assignment, respectively. In both methods, the final ACE is
estimated as the ratio of expected outcome under treatment and expected
outcome under control [cf. equation~(\ref{riskratio})].
The results of the two methods and their comparisons are presented in
the next section.

\subsection{\texorpdfstring{Results.}{Results}}
\label{3}

This section presents the results for each of the three comparisons in
terms of achieved balance and estimates of ACE.

%
%
\begin{figure}

\includegraphics{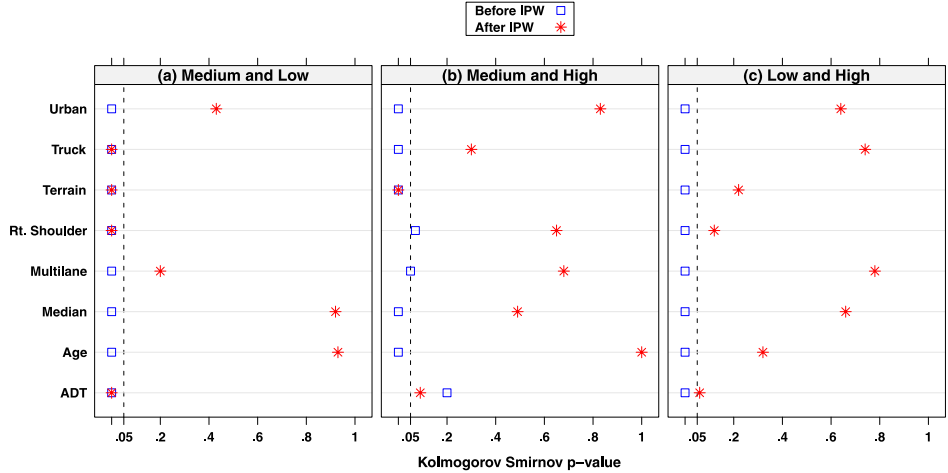}

\caption{Balance of covariates before and after weighting for \textup{(a)}
Treatment${}={}$Medium and Control${}={}$Low, \textup{(b)} Treatment${}={}$Medium and
Control${}={}$High, and \textup{(c)} Treatment${}={}$Low and Control${}={}$High. The vertical
dashed line indicates the cutoff $p$-value at 0.05 level.}
\label{fig:balance}
\end{figure}

\textit{Balance of key covariates}.
Figure~\ref{fig:balance} summarizes the effect of weighing on achieving
the balance; more detailed statistics are provided in the supplementary
materials [\citet{supp}].
The graph shows the $p$-value for the Kolmogorov--Smirnov test statistic
before and after the weighting of key covariates for each of the three
comparisons.
Figure \ref{fig:balance}(a) shows that there was a considerable
improvement in the balance after IPW for the Medium to Low comparison.
However, the variables Truck, Terrain, ADT and Right shoulder width
remain unbalanced even after weighting. Terrain was also unbalanced in
the High to Medium comparison; see Figure~\ref{fig:balance}(b). Figure
\ref{fig:balance}(c) shows that all covariates seem reasonably well
balanced for the High to Low comparison.

%
\begin{table}
\caption{Estimate of ACE based on PO framework}
\label{table:poace}
\begin{tabular}{@{}lcccc@{}}
\hline
 &\multicolumn{2}{c}{\textbf{Individual prediction}} &\multicolumn{2}{c@{}}{\textbf{Combined prediction}} \\[-5pt]
\textbf{Change in}&\multicolumn{2}{c}{\hrulefill} &\multicolumn{2}{c@{}}{\hrulefill}\\
\textbf{visibility level}& \textbf{Point estimate} & \textbf{95\% limits} & \textbf{Point estimate} & \textbf{95\% limits}\\
\hline
High to Low & 2.53 & $[2.31, 3.60]$& 3.09 & $[2.22, 4.52]$\\
Medium to Low & 1.21 &$[0.9, 1.4]$ & 1.35 & $[0.8, 1.81]$\\
High to Medium & 1.82 &$[1.67, 1.97]$ & 1.88 & $[1.5, 2.02]$\\
\hline
\end{tabular}
\end{table}

%
%

\textit{Estimation of ACE}.
The estimates of ACE for each of the comparisons are shown in Table \ref
{table:poace} for both individual and combined predictions. For the
High to Low comparison, the results from both prediction methods
suggest that application of High $\mathit{PMR}$ significantly reduces the risk
of a target crash in comparison to application of Low $\mathit{PMR}$.
Based on the results from both groups,
the risk of a target crash
on segments with Low $\mathit{PMR}$ is 3.09 times that of High $\mathit{PMR}$.
Furthermore, we can be 95 percent confident that the expected risk of a
crash on segments with low $\mathit{PMR}$ is between $2.22$ and $4.52$ times
that on segments with application of High $\mathit{PMR}$.
Similarly, based on the results from the prediction from individual
groups, the risk of a target crash on segments with Low $\mathit{PMR}$ is 2.53
times that of High $\mathit{PMR}$ with the 95 percent confidence interval [2.31,
3.60]. The ACE point estimates from the two methods are quite
different, but there is considerable overlap in the confidence
intervals, with the interval from the combined prediction being
slightly wider than from the individual prediction. This is due to
model-based extrapolation in the combined method.

The results for High to Med comparison, from both methods, also suggest
that application of High $\mathit{PMR}$ significantly reduces the risk of a
target crash in comparison to application of Med $\mathit{PMR}$, but the
expected risk is smaller in magnitude than for High to Low $\mathit{PMR}$
comparisons. The point estimates of the ACE from two methods are close
to each other with considerable overlap in the confidence intervals:
combined ACE of 1.88 (95\% CI: 1.5, 2.02) and individual ACE of 1.82
(95\% CI: 1.67, 1.97).

For the Med to Low comparison, the results indicate that there may be
no significant effect of changing the PMR level from Low to Med. The
point estimates of the ACE and the confidence intervals from the two
methods are close to each other. From the combined prediction, the risk
of a target crash on segments with Low $\mathit{PMR}$ is 1.35 (95\% CI
[0.8, 1.81]) times that of Med $\mathit{PMR}$. From the individual prediction,
the risk of a target crash on segments with Low $\mathit{PMR}$ is 1.21 (95\% CI
[0.9, 1.4]) times that of Med $\mathit{PMR}$. However, it must be noted that this
was the most difficult data subsample to attain the balance on the
covariates, and the risk ratios may be biased, due to lack of balance
over important key covariates.

\section{\texorpdfstring{Causal diagrams.}{Causal diagrams}}
\label{sec:cd}
In this section we present the Causal Diagrams (CD) framework to
estimate ACEs. We use discrete Causal Bayesian networks (CBN) to
implement the CD framework as described in Section \ref{sec:cd.intro}.
Section \ref{sec:cd.learnnetwork} briefly discusses the algorithms used
to recover a CBN from observational data, the required assumptions and
estimation procedures for the ACE.
Section \ref{sec:cd.results} presents the results of the analysis.

\subsection{\texorpdfstring{Causal diagrams and components of a causal model.}{Causal diagrams and components of a causal model}}
\label{sec:cd.intro}
In the CD setting, a causal model is used as the fundamental element to
estimate causal effects, in contrast with the PO model, where potential
outcomes are the fundamental quantities.
Let $V$ denote the set of variables representing the attributes of a
road segment which includes both the treatment assigned to a segment
and its safety outcome. A~Causal Model describes the causal relations
(in the form of conditional independence) among the variables in~$V$.
The qualitative part of the model is represented by a graph using a set
of nodes and edges, and the quantitative part by a set of conditional
probability distributions associated with each node in the graph.

In our analysis, we represent the Causal Model by using a discrete
Causal Bayesian Network (CBN) for implementing the CD
framework.\footnote{CBNs with continuous variables are possible, but
algorithms for handling arbitrary continuous distributions are not well
developed. Also, many algorithms cannot handle mixed BNs (mixed here
refers to the combination of continuous and discrete variables) that
have continuous parents of discrete children.} A CBN consists of a
directed acyclic graph (DAG) and a set of probability distributions
associated with each node, represented by a conditional probability
table (CPT). Figure \ref{gx} shows an example of such a graph, where,
for instance, Safety ($S$) is a \textit{child} (\textit{ch})  of $\mathit{ADT}$ and $\mathit{PMR}$,
and, thus, these are its \textit{parents} (\textit{pa}). For more details on graphs
and graphical models, see \citet{Lauritzen99causalinference}. The
discrete versions of the variables as defined in Table \ref{define}
were used.

The problem of causal inference involves learning the causal structure,
represented by a DAG and a CPT, from data. The ACE of a treatment under
intervention is estimated using \textit{intervention theory}, as explained
in the next section.

\subsubsection{\texorpdfstring{Causal diagrams as models of intervention.}{Causal diagrams as models of intervention}}
According to \citet{Pearl00causalitymodels}, CBNs can be regarded as
models of interventions if it is assumed that a DAG models the causal
mechanism which generated the data. (See Section~\ref
{sec:cd.learnnetwork} for a review of this assumption.)

\begin{figure}

\includegraphics{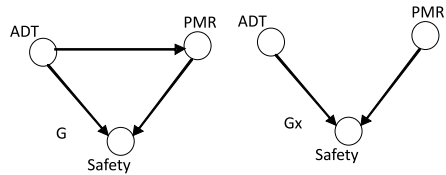}

\caption{An example of the interventional distribution. The graph $G$
represents the original DAG. The mutilated graph $G_{x}$ under the
intervention of forcing $\mathit{PMR}$ to take a particular value is obtained by
deleting the arcs between $\mathit{PMR}$ and its parents
(e.g., $\mathit{ADT}$).}
\label{gx}
\end{figure}

Under the above assumption, the edges in a DAG are used to specify the
changes in the joint distribution of variables $V$ due to external
intervention. For instance, in Figure \ref{gx}, forcing the node $\mathit{PMR}$
to take a particular value, say, $\mathit{Low}$, amounts to lifting the existing
mechanism on $\mathit{PMR}$ and putting it under the influence of a new
mechanism whose action is to force $\mathit{PMR}$ to the value $\mathit{Low}$, keeping
everything else constant. This action is mathematically represented by
$\mathit{do}(\mathit{PMR}= \mathit{Low})$. The effect of ``setting'' a node to a fixed value
corresponds to applying the low $\mathit{PMR}$ treatment to all the segments in
the sample.
Such interventions are modeled in a DAG $G$ by creating a new mutilated
DAG $G_{\mathit{PMR}}$ from $G$. In $G_{\mathit{PMR}}$, the links between $\mathit{PMR}$ and its
parents are removed, keeping the rest of the graph the same. The
distribution imposed by the new graph $G_{\mathit{PMR}}$ under the condition
$\mathit{PMR}= \mathit{Low}$ represents the effect of intervention and is called the
post-intervention distribution; for example, see Figure \ref{gx}.

\subsubsection{\texorpdfstring{Causal effect and ACE.}{Causal effect and ACE}}\label{sec5.1.2}
Given the safety outcome $S$ and the\break treatment variable $\mathit{PMR}$, the
causal effect of $\mathit{PMR}$ on $S$, denoted by\break $P[S|\mathit{do}(\mathit{PMR} = i)]$, where $i
\in\{\mathit{Low}, \mathit{Med}, \mathit{High}\}$, is a function from $\mathit{PMR}$ to the space of
probability distribution on $S$. For each realization of $\mathit{PMR}$,
$P[S|\mathit{do}(\mathit{PMR} = i)]$ gives the probability of $S= s$ induced from the
mutilated graph $G_{\mathit{PMR}}$ and substituting the value of $\mathit{PMR}$ as $i$ in
this graph.

Given a causal diagram in which all the parents of manipulated
variables are observed, the causal effect can be estimated from passive
or noninterventional data. However, when some parents of a child node
$ch$ are not observed, $P(ch|pa_i)$ may not be estimable in all cases.
A graphical test has been provided by \citet{Pearl00causalitymodels}, Chapter 3, to find out when $P[S|\mathit{do}(\mathit{PMR})]$ is
estimable from the observed data. In the present case, we make the
assumption that all potential confounders are included in the analysis.
This is a strong assumption and must come from subject matter experts.
These assumptions are reviewed in Section \ref{sec:cd.learnnetwork}.

When $\mathit{PMR}$ has two possible states ($\mathit{Low}$ and $\mathit{Med}$), the ACE is given
by the following equation:
%
\begin{equation}\label{causaleffect}
\mathit{ACE}_{\mathit{Med}\mathrm{\ to\ }\mathit{Low}} = \frac{P(S = 1|\mathit{do}(\mathit{PMR}=\mathit{Low}))}{P(S = 1|\mathit{do}(\mathit{PMR}=\mathit{Med}))},
\end{equation}
where $P[S = 1|\mathit{do}(\mathit{PMR}=\mathit{Low})]$ is the marginal probability in
$G_{\mathit{PMR}}$\footnote{Conditioning on so-called colliders can actually
introduce bias in the ACE; see \citet{Pearlci}. In simple terms, a
variable is a collider if it has two arrows into it. In the present
case, there are no such colliders on the path between Safety and
$\mathit{PMR}$.} of $S= 1$ under the intervention $\mathit{PMR} = \mathit{Low}$; similar
expressions are used for the other two comparisons, \textit{High} to \textit{Low} and \textit{High} to \textit{Med}. Also, notice the similarity to
equation (\ref{riskratio}) from the PO framework.
In the next section we describe the algorithms that were used to learn
the components of the CBN.

\subsection{\texorpdfstring{Learning CBNs from data and estimation of ACE.}{Learning CBNs from data and estimation of ACE}}
\label{sec:cd.learnnetwork}
Structure learning algorithms are used to recover a DAG $G$ and
parameter learning algorithms are used to estimate the CPTs, which then
lead to estimation of ACE.\looseness=-1

\subsubsection{\texorpdfstring{Structure learning.}{Structure learning}}
\label{assump}
Learning the structure of causal networks from observational data has
received a thorough treatment in the literature; see
\citet{Pearl91atheory}, \citet{Heckerman96atutorial}, \citet{PeterSpirtes1991}. The most
common strategies fall into two different classes called \textit{constraint based learning} and \textit{score based learning}. We adopt a
simple combination of both approaches to learn the structure of the
CBN. Our approach is similar in principle to \citet{bnsearch}. The PC
algorithm [\citet{causationprediction}] is used as a constraint based
strategy to recover a DAG from the data. This DAG is supplied as an
initial input to the score based learning strategy, which then attempts
to find an optimum DAG. The simulated annealing strategy of \citet
{banjo} and the scoring function proposed by \citet
{Heckerman94learningbayesian} are used to search for optimum scored
CBN. The scoring search is implemented in Java using the BANJO library
[\citet{banjo}] and the constraint search is performed using the BNT
toolbox in Matlab [\citet{bnt}]. Since the scoring method need not
produce the globally optimum structure of the CBN, we used the 10 best
networks recovered by the algorithm and performed Bayesian model
averaging to estimate the ACE. For details on the averaging, refer to
\citet{rafterygraphical}, \citet{hoetingbma} and \citet
{Heckerman94learningbayesian}.

Irrespective of the strategy used, a DAG can be recovered from
observational data, up to  $d$-separation equivalence [\citet
{Pearl00causalitymodels}, Chapter 1], only if the three assumptions
outlined below are satisfied. Causal interpretation of CBN is possible
because of these assumptions, which are in general untestable from
observational data and must come from subject matter experts.

\textit{Causal Markov Assumption}:
The Causal Markov Assumption (CMA) states that given the values of a
variable's immediate causes (i.e., its parents), the variable is
independent of its nondescendants [\citet{Pearl00causalitymodels},
Chapter 1]. This assumption implies that we must include in the model
every variable that is a cause of two or more other variables. It also
implies Reichenbach's [\citet{Reichenbach}] common cause assumption,
which states that, if any two variables are dependent, then one is a
cause of the other or there is a third variable causing both.

The natural question that arises is what are the immediate factors that
affect the safety of a segment? For instance, is driving at a high
speed considered an immediate cause of reduced safety? To understand
the CMA in light of safety, we need to consider factors that can cause
a crash.
These factors can be divided into three broad categories: road user
(driver), the vehicle, and roadway characteristics (environmental
conditions, roadway volume, etc.).
Generally, information on the factors related to drivers and the
vehicle is available only for vehicles involved in a crash, and not for
noncrash vehicles. Thus, in a driver level analysis, most of the data
would be missing. Also, the immediate causes of crashes (75 percent of
which are due to human error [\citet{humanerror}]) become very specific
to a particular crash and are governed by complex human behavior which
is difficult to model and predict. To avoid these issues, analysis is
done at the segment level. Only stable attributes of a roadway segment
are included in the analysis; specific human factors are included in
the error terms considered to be stochastic in nature. Thus, CMA is
treated as a guiding principle rather than an assumption, where it
defines the granularity of the model being considered, ensuring that
all relevant causes, as defined by subject matter experts and past
experiments, are included in the analysis.

\textit{Faithfulness}: The faithfulness assumption ensures that the
population that generated the DAG has exactly those independence
relations specified by the DAG structure and no additional
independencies. If there are any independence relations in the
population that are not a consequence of the Causal Markov condition,
then the population is unfaithful. By assuming Faithfulness, we
eliminate all such cases from consideration.\footnote{This assumption is
controversial; see the discussion section.}

\textit{Latent variables}: This assumption states that there are no
hidden variables in the model that violate the causal Markov condition.
That is, all of the variables that effect more than two variables in
the model are observed and included in the database. Again, this is a
strong assumption, whose validity could be ensured by verification from
subject matter experts. For instance, the definition of safety ensures
that the causes due to driver and weather factors do not influence the
outcome, or else these would have to be entered into the model as
latent variables.

\subsubsection{\texorpdfstring{Parameter learning.}{Parameter learning}}\label{sec5.2.2}
The parameters of the CPT are modeled using \textit{Dirichlet
distributions} and the usual assumptions of parameter independence are made.
For details on parameter learning, see \citet
{Heckerman94learningbayesian}. The Bayesian Dirichlet Equivalent
Uniform Priors (BDEU) were used to compute the parameters of the CBN.

The Dirichlet hyper parameters $\alpha_{x_i,\pi_i}$ are specified by
the following equation:
%
\begin{equation}
\alpha_{x_i,\pi_i} = \alpha\times p(x_i,\pi_i ),
\end{equation}
where $\alpha_{x_i,\pi_i}$ pertains to variable $X_i$ in a state $x_i$
given that its parents are in joint state $\pi_i,$ for $i=1,\ldots ,n$,
where $\alpha$ is the number of pseudo-counts, and $p$ is a (marginal)
prior distribution of pseudo-counts; this ensures the
likelihood-equivalence of Markov equivalent structures [\citet
{Heckerman94learningbayesian}]. The value of $\alpha$ is taken to be 1.
The distribution~$p$ is chosen to be uniform between 0 and 1 for all
variables (representing noninformative prior), that is, for any CPT,
each parent-child combination is given an equal probability.

\subsubsection{\texorpdfstring{Estimation of ACE.}{Estimation of ACE}}
\label{sec:cd.ace}
There are several methods in the literature [\citet{infmethod}] to
efficiently perform inference in a CBN.
We computed the marginal probability of $S $ by using the \textit{junction
tree algorithm} that performs exact inference. Recall that the ACE of
$\mathit{PMR}$ on $S$ is estimated as the ratio of the expected value of safety
under the intervention level corresponding to the treatment and the
expected value of safety under the intervention level corresponding to
control [cf. equation~(\ref{causaleffect})].
A full Bayes model was specified and the confidence in the value of the
ACE was estimated by computing a~$95$ percent Bayesian credible interval.

\subsection{\texorpdfstring{Results.}{Results}}
\label{sec:cd.results}
\begin{figure}

\includegraphics{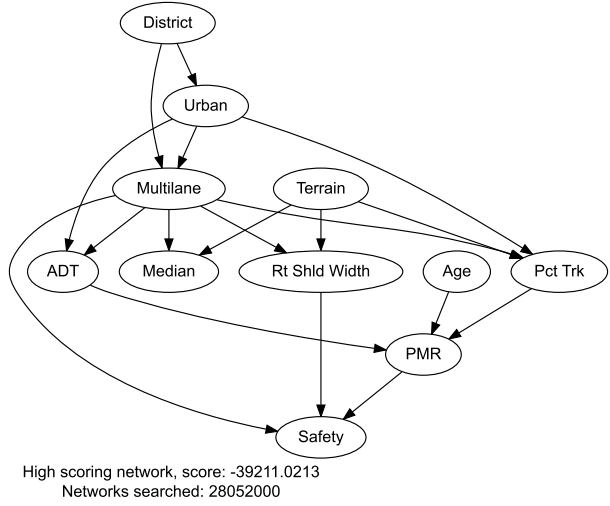}

\caption{The best scoring DAG recovered by the search algorithm.}
\label{bestdag}
\end{figure}

The DAG with the highest score is shown in Figure \ref{bestdag}. Notice
that there is no $\mathit{Low}$  $\mathit{speed}$ variable in this DAG. This could be
because given the combination of variables like $\mathit{ADT}$, $\mathit{Median}$ and
$\mathit{Multilane}$, the value of $\mathit{Low}$  $\mathit{speed}$ is completely determined, and
the discretization of $\mathit{ADT}$ into two levels makes it highly collinear
with $\mathit{Low}$  $\mathit{speed}$.
%
%
It was surprising to see the safety of a segment directly unaffected by
$\mathit{ADT}$ in this particular DAG, since it is commonly observed that the
higher the $\mathit{ADT}$, the higher the probability of a target crash on a
segment. However, two of the top 10 graphs show that $\mathit{ADT}$ does indeed
affect safety. A possible reason could again be the discretized $\mathit{ADT}$
variable, which is also highly correlated with the $\mathit{Multilane}$
variable; segments with more than two lanes generally have high $\mathit{ADT}$.
Similar problems were encountered in the PO framework. This could be
the reason why the $\mathit{Multilane}$ indicator affects safety in 8 out of the
10 highest scoring models.

%
\begin{figure}[b]

\includegraphics{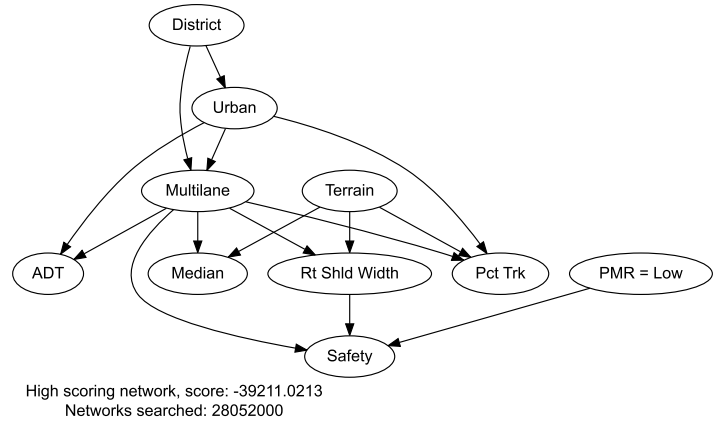}

\caption{The mutilated graph under the manipulation of PMR${}={}$Low.}
\label{mudag}
\end{figure}

Figure \ref{mudag} shows the mutilated DAG used to model the effect of
intervention on the $\mathit{PMR}$ levels. As noted earlier, the mutilated DAG
is formed by deleting all the edges from the original DAG that direct
into the $\mathit{PMR}$ variable, and fixing the value of $\mathit{PMR}$ at a particular
level. The marginal probability distribution of safety in such a DAG
represents the effect of manipulating the $\mathit{PMR}$ variable on $S$. We
computed the full Bayesian posterior of ACE using Monte Carlo
simulation, averaging over the 10 best selected networks.\looseness=1

\begin{table}
\caption{Estimate of ACE based on CD framework}
\label{causalace}
\begin{tabular}{@{}lcc@{}}
\hline
\textbf{Change in visibility level} & \textbf{Point estimate} & \textbf{95\% limits}\\
\hline
High to Low & 3.12 & $[2.32, 4.11]$\\
Medium to Low & 1.79 & $[1.31, 2.28]$\\
High to Medium & 1.86 & $[1.60, 2.17]$\\
\hline
\end{tabular}
\end{table}

\begin{figure}[b]

\includegraphics{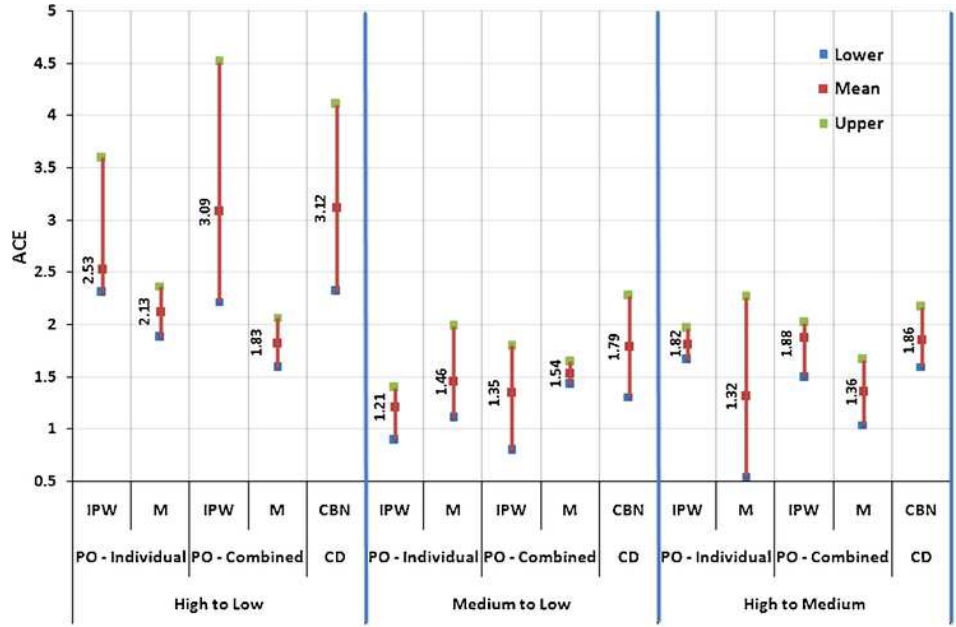}

\caption{Overlap between the confidence intervals of ACE from both frameworks.}
\label{cioverlap}
\end{figure}

Table \ref{causalace} shows the final results of the effect of $\mathit{PMR}$ on
safety, computed via the Causal Diagrams approach. The results suggest
that higher $\mathit{PMR}$ levels correspond to significantly lower risk of a
target crash on all comparisons. In particular, the risk of a target
crash on a segment with Low $\mathit{PMR}$ is 3.12 times that of High $\mathit{PMR}$ with
a 95\% CI of [2.32, 4.11] and 1.79 times that of Med $\mathit{PMR}$ with a 95\% CI of [1.31, 2.28]. Similarly, the risk of a target crash on a
segment with Med $\mathit{PMR}$ is 1.86 times that of High $\mathit{PMR}$ with a~95\% CI
of [1.60, 2.17].

\section{\texorpdfstring{Comparison of results and discussion.}{Comparison of results and discussion}}
\label{sec:comp.res}
This section compares and discusses the analyses and results from the
PO and CD frameworks. In addition, it also includes the results from
the popular PO alternative to IPW, the propensity score matching, whose
implementation details are available in the supplementary document
[\citet{supp}].

\subsection{\texorpdfstring{Comparison of results and implications.}{Comparison of results and implications}}
\label{compare}
Figure \ref{cioverlap} shows the point estimates and confidence
intervals of ACEs from both frameworks. Specifically, it shows the
overlap among the estimates of ACEs from the following methods:
Combined and Individual approaches, using Inverse Propensity Score
weighting (IPW) and using Propensity Score Matching (M) with regression
adjustment from the PO framework, and Causal Bayesian Networks (CBN)
from the CD framework. There are a couple of important points to be noted.

In terms of a general trend, the results of all methods are consistent
with each other; they show that increased $\mathit{PMR}$ levels generally lead
to a reduced or unchanged risk of a target crash which agrees with
engineering expectations. In terms of magnitude of the effect, however,
there is a noticeable and in some cases statistically significant
difference across different implementations. The CBN point estimates
are consistently higher than both the IPW and M point estimates.

In general, the 95\% confidence regions from the three methods show
significant overlap across the three comparisons, and thus lead to the
same conclusions in terms of the expected strength of the causal
effect. The most consistent result, statistically and with respect to
transportation engineering logic, is for the High to Low comparison
where all three methods indicate that there is a significant reduction
in risk with application of High $\mathit{PMR}$ in comparison to Low $\mathit{PMR}$ even
though the matching results are significantly lower in magnitude, in
particular, for the PO-combined M estimate. The IPW and CBN results
display strong correspondence with the exception of Med to Low
comparison. The inference based on IPW results implies that the risk of
a target crash is not significantly lower on segments with the
application of Med $\mathit{PMR}$ levels when compared to Low $\mathit{PMR}$ levels,
whereas both CBN and M based confidence intervals imply statistically
significant risk reduction and are more in line with engineering intuition.

The most curious result with respect to engineering expectation occurs
for the High to Med comparison. Here, it was expected that there would
be a small or no statistically significant effect. Only the matching
results support this assertion, and, in particular, only the
PO-individual M result claims no effect. The PO-combined M estimates
are closer to IPW and CBN results, with the latter two having a very
strong correspondence that implies a significant reduction in safety
risk with the application of High PMR compared to Med PMR. It should be
noted, however, that the PO-individual M estimate has the highest
variability, with a significant CI overlap with other estimated
effects, thus, it may be difficult to render a solid practical decision
based solely on statistical significance or lack thereof.
Alternatively, one can argue that PO-combined, IPW and the CBN
estimates are based on extrapolation when compared to the PO-individual
M estimates due to differences in data subsamples being used. Further
examination of the results is needed, which is discussed below.

\subsection{\texorpdfstring{Discussion of results.}{Discussion of results}}
\label{discuss}

There could be several reasons for the differences in the results from
the PO framework and the CD framework. One is distributional support.
The variables used in the CBN setting were discretized to ensure the
use of efficient algorithms. On the other hand, there was little
discretization performed in the PO model. Specifically, the variable
``age'' was used as a continuous variable in the IPW estimation, and
``age,'' ``Rt. Shoulder width,'' ``Percentage Trucks'' and ``ADT'' were
used as continuous variables in the Matching estimation.\footnote{These
choices were based on two (conflicting) policies, the first was to
ensure similarity to the variables used in the CBN setting, and the
second to ensure balance, the second one being given more priority.}

As noted, the CBN and combined-IPW results are close to each other.
This is due to the similarity in empirical estimation of ACE. In both
frameworks, causal effect is defined as a contrast between the outcomes
of the \textit{same} unit with and without treatment. However, to avoid
the problem of missing potential outcomes, the individual IPW estimator
compares \textit{similar} units, whereas the combined estimator compares
the \textit{same} units. In the latter case, the problem of unobservable
potential outcomes is avoided by using the outcome model for
prediction. This is similar to applying the ``do'' operator to all
units and contrasting the outcomes under different
manipulations.\looseness=1

The differences between results of matching and results of CBN and IPW
are due to differences in the data set, more specifically, due to loss
of data in matching. To ensure overlap and balance, matching discards data.
\citet{dehejiawahba1998} and \citet{heckman1998matching} point out
that without overlap the results would be sensitive to the
specification of either the propensity score or outcome model. While
our matching procedure attained both balance and sufficient overlap for
all comparisons, there was a~significant loss of observations; see the
supplementary document for details. For instance, for High to Med case,
matching discards 83\% of the data. Thus, the PO-individual M estimate
of no significant change for High to Med may not be representative of
the whole population. On the other hand, matching has the smallest loss
of information for Med to Low case. The results from the CBN and
matching show that there could be some significant effect in changing
the $\mathit{PMR}$ level from Med to Low, whereas the IPW results indicate that
there may not be a significant effect. However, the IPW results may be
biased because of poor balance achieved in the data for this comparison
(cf. Figure~\ref{fig:balance}).

Some of the differences between IPW and CBN could also be due to the
use of Bayesian estimation in CBN. For instance, in CBN, Bayesian model
averaging was used to account for the possibility of recovering a local
minima. Moreover, all of the variables in CBN were discrete, which lead
to large and sparse CPTs. The estimates from the PO framework were also
affected by sparseness, although in part mitigated by the presence of
continuous variables.

\subsection{Which method to use?}\label{sec6.3}
The choice of method is greatly influenced by the assumptions made. No
method can be completely assumption free, in fact, all causal
inferences (from observational data) must be based on causal
assumptions. The major causal assumptions in each framework were the
Causal Markov Assumption (CMA) in CD and the unconfoundedness
assumption in the PO framework. We conjecture that CMA is a stronger
assumption than unconfoundedness, as it pertains to all the variables
in the problem at hand, whereas the latter relates to the treatment
variable and the potential outcomes. Furthermore, recent work has shown
that unconfoundedness alone may not be sufficient to identify
appropriate covariates for inclusion in the propensity score model; see
\citeauthor{Pearl00causalitymodels}   (\citeyear{Pearl00causalitymodels,pearl2009class}).

The PO framework requires overlap and balance assumptions, whereas the
CD framework does not. In the broadest sense, the balance assumption
ensures that there are no-pretreatment differences between the groups
being compared, and the overlap assumption ensures that the estimates
do not rely too much on the functional specification of the model.
However, in the case of CD, the (qualitative) causal model is assumed
to be known completely.\footnote{This may not be always the case,
especially in the social sciences.} Moreover, the framework does not
require any explicit functional specification of relation between the
variables, given the CMA [and some additional assumptions; see \citet{Pearl00causalitymodels}, Chapter 3, for technical details and
mathematical proofs]. Based on this discussion, we suggest the
following guidelines for deciding which method to use.

If little is known about the data generating process or the causal
mechanism, the analyst should go as ``nonparametric'' as possible. For
example, one could use matching (preferably by specifying the
propensity score model using a nonparametric estimator such as GBM),
ensure sufficient overlap, and compute average causal effect from the
observed data by using individual prediction. However, this may depend
on the specification of the propensity score model, and, more
importantly, in the case of significant loss of data (as in our case),
it is not clear as to which subpopulation the ACE estimates apply. It
must also be noted that this strategy may not guard against the
inclusion of inappropriate covariates in the propensity score model
(such as colliders and bias amplifying covariates).

One could attempt to recover the data generating process from
observational data under further assumptions of faithfulness, combined
with partial expert knowledge. However, \citet{robins1999impossibility}
show that when the probability that variables in the causal model have
no common unobserved causes is small relative to the sample size,
analysis carried out using faithfulness can lead to inappropriate conclusions.

On the other hand, if the data generating mechanism is known (even
qualitatively), the mechanism can be summarized in the form of a causal
diagram. The causal diagram may incorporate the mechanism related to
treatment assignment and/or the response to the treatment. Such a
causal model can be used as a guide to estimate the propensity score
model as well as the outcome model (which can then be used with other
adjustment methods such as weighting, matching, etc). The dependence on
the functional form between variables can be reduced by using
categorical variables\footnote{This may come with a set of its own
problems, for example, sparseness.} and/or by using nonparametric
estimators such as GBM, as in our case.

In a real data setting, it is always better to compare the results from
different methods. In the present study, it is clear that for different
comparisons of PMR levels, different methods show consistency based on
which assumptions are being violated. For instance, in the High to Low
$\mathit{PMR}$ case, all methods show good agreement. In the Med to Low case,
Matching and CBN show good agreement (IPW results may be biased due to
lack of balance). In the Med to High case, IPW and CBN show good
agreement (Matching may be biased due to significant loss of data).

The CBN results could also be biased if the causal model recovered by
the data is not close to the truth. However, there is evidence that the
CBN may be less biased when compared to other methods. In the
comparisons where the data are well balanced (High to Low and High to
Med) and there is considerable overlap (High to Medium), the CBN
results are in close agreement with the IPW, which indicates that the
model may be close to the truth. Since the true ACE is unknown, the
only test for validity of the results is by implicit agreement of
results from different methods. If different methods provide the same
answer, the answer must be close to the truth, or, in the worst case,
all methods fail to capture the same aspect of the true
model.\looseness=-1

\subsection{\texorpdfstring{Future work.}{Future work}}\label{sec6.4}
To obtain a better comparison of the methods, future studies should aim
at using data from simulation. The true causal effect of the population
would be known a priori, and the quality and size of data can be
controlled. Other advances on this exploratory work can be made by
using more complex causal modeling methods. For instance,
discretization of the $\mathit{PMR}$ treatment variable can be avoided by using
a dose-response model [\citet{dose}]. In the current study,
temporal/spatial correlations may exist, though evidence was not found in
these data. The $\mathit{PMR}$ treatment can be modeled as a time varying
treatment [\citet{timevary}] to take into account such correlations.
Specification of the assignment mechanism for the $\mathit{PMR}$ treatment
variable is convenient when compared to other possible countermeasures,
such as roadway lighting. The assignment mechanism for lighting is
generally influenced by factors such as local design policies,
complaints from residents and may also be related to past crash
history. Such assignment mechanisms may prove difficult to model and
may require the use of latent variables.
Also in the current study, we did not explicitly consider uncertainty
in the imputed (using ANN) PMR levels and uncertainty in the
measurement process; rather a mean estimate was used. Both of these are
important issues that should be carefully considered as part of future work.
Sampling zeros were encountered in both the PO model as well as the CBN
setting. In the PO framework, such sampling zeros created problems in
achieving balance over interactions of covariates (specifically in the
matching estimator). In the CBN setting, the use of Bayesian Inference
in part addressed this problem. A similar approach in the PO framework
would be to use a full Bayes model of both the propensity scores as
well as safety outcomes [\citet{rubin2008}]. These explorations are
left to the scope of future work.

\section{\texorpdfstring{Conclusion.}{Conclusion}}
\label{con.dis}
The examination of causal inference methods to transportation safety
data reveals that there is considerable scope of their application to
estimate safety effects of a countermeasure.
A comparison was made between the PO framework and the CD framework.
More specifically, the results based on three different implementations
of these frameworks on a real data set were compared: Inverse
Propensity Score Weighting with regression adjustment and Propensity
Score Matching with regression adjustment versus Causal Bayesian
Network.

Although the general trend of results seem to be consistent, we found
that the magnitude of ACEs are sensitive to the method used and to the
assumptions being violated.
In real data sets, it is very likely that some assumptions will be
violated. Depending upon which assumptions are appropriate,
different methods should be used. Assumptions should be considered a
priori. If possible, the analyst should run multiple implementations to
compare the results for consistency.
In conclusion, we suggest the use of the PO framework supplemented by a
qualitative causal diagram as a rich framework to estimate the safety
effects of countermeasures in transportation studies.

\section*{\texorpdfstring{Acknowledgments.}{Acknowledgments}}
The authors are grateful to the Editor, Associate Editor and an
anonymous referee whose
comments have greatly improved the content of this work. We are also
thankful to the
North Carolina Department of Transportation for providing the data, and
Dr. Kenneth Opiela from the Federal Highway Administration Office of
Safety R\&D for
coordinating with the North Carolina Department of Transportation to
obtain the data.

\begin{supplement}
\stitle{Supplement to ``Causal
inference in transportation safety studies: Comparison of potential
outcomes and causal diagrams''\\}
\slink[doi,text={10.1214/10-}]{10.1214/10-AOAS440SUPP} 
\slink[doi,text={AOAS440SUPP}]{10.1214/10-AOAS440SUPP}
\slink[url]{http://lib.stat.cmu.edu/aoas/440/supplement.pdf}
\sdatatype{.pdf}
\sdescription{This document contains additional details about the
Matching and Inverse Propensity
score estimators and the top ten graphs recovered by the graph learning
algorithm.}
\end{supplement}


\printaddresses

\end{document}